\documentclass[twocolumn,10pt]{revtex4-2}

\begin{document}

\newcommand{\be}{\begin{equation}}
\newcommand{\ee}{\end{equation}}
\newcommand{\bea}{\begin{eqnarray}}
\newcommand{\eea}{\end{eqnarray}}

\title{Strange Sounds}

\author{D. V. Khveshchenko} 
\affiliation{Department of Physics and Astronomy, 
University of North Carolina, Chapel Hill, NC 27599}

\begin{abstract}
\noindent
This note addresses the effects of long-ranged  
and/or retarded interactions on the bosonic collective modes in the so-called 'strange metals'.
Recently, there have been conflicting reports on the very existence 
of such stable collective excitations and their properties. 
Extending a number of approaches that were previously used in the analyses of the standard 
Fermi liquid one finds
evidence for, both, conventional and novel behaviors 
of the non-Fermi-liquid counterparts of the zero-, shear-, and other 'sounds'.  

\end{abstract}

\maketitle

{\it Strange metals}\\

There has been a strong  
interest in the hydrodynamic behavior of electronic systems
and some evidence in support of this interaction-dominated regime was  
reported in a number of materials, including graphene, $GaAs$, $PdCoO_2$, etc. 

The experimental findings suggestive of such behavior also bolstered 
the attempts to discover hydrodynamic
features in the so-called 'strange metals' which moniker is 
often used in reference to the generic non-Fermi-liquids (NFL)
states of compressible fermion matter. 

Despite the decades of intensive studies, a systematic
classification of the NFL is still missing 
and such states are characterized largely by what they are not - namely, not the ordinary Fermi liquids (FL).
Therefore, one potential NFL classification could be constructed on the basis of spectroscopy  
of their bosonic collective excitations or 'sounds'. Their spectra  can be extracted from the acoustic density response, 
electromagnetic polarization, optical conductivity, etc. 

Many of, both, established and conjectured NFL systems fall within the broad category of 
(non-)relativistic fermions coupled via an overdamped bosonic field with the propagator
\be 
D(\omega,q)={1\over |\omega|/q^{\alpha}+q^{\beta}}
\ee  
Its non-relativistic (finite density) variant has been encountered in a whole variety of condensed matter problems: 
itinerant ferro- and anti-ferro- magnets \cite{hertz}, 
electromagnetic response in ordinary metals and plasmas \cite{reizer}, 
spinon gauge theories of spin liquids \cite{gaugeold,stampdvk,chubdvk},  
compressible QHE \cite{qhe}, Ising nematics and other 
examples of Pomeranchuk instabilities and Lifshitz transitions \cite{chub}.  
This topic has also been addressed in the nuclear and high-energy studies 
where the mode in question is a genuine 
(non-)abelian gauge field \cite{pethick}.  

In $d$ spatial dimensions and to first order in its strength, the interaction 
(1) endows the fermions with a singular and largely momentum-independent self-energy
\be 
\Sigma(\omega)\sim \omega^{1/z}~~~~~~~z={d-1+\alpha\over {\alpha+\beta}}
\ee 
Despite having been around for quite a while (and contrary to the occasional claims to the opposite \cite{gaugenew}), 
this fundamental problem is still waiting for its satisfactory solution.       
In particular, the ultimate form of the fermion propagator - especially, at finite temperatures
and/or in the case of unbroken gauge symmetry - 
is yet to be ascertained (see \cite{mother} and references therein).\\ 

{\it Collective modes of interacting fermions}\\

While many of the questions pertaining to the one-fermion 
properties in the presence of singular interactions 
remain unanswered, there has long existed a certain consensus regarding the two-particle 
ones. Namely, at low transferred momenta and frequencies the 
density and current response functions are normally expected to 
manifest, by and large, the ordinary FL behavior \cite{qbe}.  

A recent follow-up of the original work \cite{qbe} addressed the issue 
of a possible (non)existence of the undamped zero- and other 
sound modes in the presence of the interactions (1).
In such work, this assertion was either supported \cite{mandal1}, refuted \cite{else,mandal2}, 
or found to depend on the interaction strength  \cite{deban}.

Specifically, in Ref.\cite{deban} it was argued that whether or not a zero sound mode 
can survive as an undamped excitation depends on the competition between the (nearly instantaneous) 
FL-like scattering processes with large momentum
transfers  and those with small momenta that are governed 
by the singularly frequency-dependent (retarded) interaction 
responsible for the NFL behavior.
Generically, one could then expect
the undamped zero-sound mode to possibly exist in the weak coupling limit, whereas at strong couplings
it would be buried inside the particle-hole continuum. 

The above conclusions were drawn from the different analyses of the quantum Boltzmann equation (QBE). 
To that end, it has long been  
argued that the latter can be used even in the absence of well-defined Landau-Fermi quasiparticles,
provided that the single-particle Green function remains a sharp function of 
momentum around the Fermi value $k_F$.

The standard derivation of the QBE makes use of the Keldysh 'lesser' function 
which reduces to the generalized distribution function 
\be
f(\epsilon,\theta | t,{\bf r})=-i\int {d\xi_p\over 2\pi} G_{<}(\epsilon, {\bf p} |t, {\bf r})
\ee
upon integrating over the quasiparticle energy
$\xi_p\approx vp_{\parallel}$ proportional to the momentum
normal to the fiducial Fermi surface (FS) and measured with respect to $k_F$. 
Hereafter $\theta$ 
is the angle (in the 2d case, the only one) parametrizing the continuous FS
and $v$ is the Fermi velocity in a generic ('non-flat') electronic band. 
  
By further integrating over the frequency one obtains an angular-resolved variation of the chemical potential 
\be
\rho(\theta | t, {\bf r})=\int {d\epsilon\over 2\pi} f(\epsilon,\theta |t, {\bf r})
\ee
which represents a local FS displacement and can be expanded over the space-time Fourier components  $\rho(\theta |\omega, {\bf q})$.  

Alternatively, a pertinent kinetic equation can be obtained from the Bethe-Salpeter one or, else, 
from the equivalent eigenvalue problem for the two-particle scattering vertex.
Its analysis then yields spectra of the particle-hole excitations, including their bound states (if any)  
which correspond to the collective modes. 

The standard form of the kinetic equation in the presence of an external electric field $E$ reads 
\bea
(\omega-vq\cos\theta)\rho(\theta)-
\int {d\theta^{\prime}\over 2\pi}F(\theta-\theta^{\prime}|\omega)
(\rho(\theta)-\rho(\theta^{\prime}))=\nonumber\\ 
=I[\rho(\theta)]+ vE\cos\theta~~~~~~~~
\eea
where  the quasiparticle interaction function $F(\theta |\omega)$
is determined by the real part of the same integral kernel 
 whose imaginary part yields the collision integral $I[\rho(\theta)]$.
For one, the singular coupling (1) gives rise to the typical energy transfer $\omega$ which is small compared 
to the momentum one, $q\approx q_{\perp}\sim \omega^{1/z}$ (hereafter, in the absence of tuning and other competing parameters all the frequencies are measured in units of the Fermi energy of order $vk_F$). 

For comparison, the earlier work of Refs.\cite{mandal1,else,mandal2} focused on the collisionless regime.
Indeed, in several relevant contexts the interaction (1) would be mediated by a physical boson, and
 neglecting the collision integral would be in line 
with the assumption that a rapidly thermalizing boson subsystem quickly attains equilibrium, thus  
ceasing to act as a momentum sink for the fermions.

Conversely, if the bosons were out of equilibrium, 
they would no longer serve as a heat bath for the fermions, and the energy conservation would need  
to be applied to the entire fermion-boson system, thus nullifying its overall collision integral. 
By introducing an additional bath (e.g., phonons), one could shift the (otherwise, vanishing)
eigenvalue of the corresponding hybrid mode away from zero, though. In what follows,
no analysis of this complex situation will be attempted.  

Also, under the condition $\omega<<vq$ the interaction-induced
scattering becomes quasi-elastic, thus allowing one to approximately linearize
the collision integral, akin to the case of static disorder. Next, expanding over the angular harmonics 
one can convert Eq.(5) into the system of coupled equations 
\bea
(\omega+i\gamma_l)\rho_l(\omega,q)=~~~~~~~~~~~~\nonumber\\
={v\over 2}\sum_{\pm}(q(1+F_0-F_l)\rho_{l\pm 1}(\omega,q)+E\delta_{l,\pm 1})~~~~~~~~~~
\eea
where $\gamma_l$ and $F_l$ are the $l^{th}$ the angular harmonics of the linearized collision integral
and the Landau function.

Alternatively, the kinetic equation can be cast in the form obtained in the original work of \cite{qbe} 
\bea
(\omega(1+F_0-F_l)+i\gamma_l)\rho_l(\omega,q)=\nonumber\\
={v\over 2}
\sum_{\pm}(q\rho_{l\pm 1}(\omega,q)+E\delta_{l,\pm 1})
\eea  
Either of the recursion relations (6) and (7) can then be viewed as a discretized Schroedinger 
equation on the 1d chain whose sites are labeled by $l$
\cite{sarma}. 
Technically, the $l$-dependent hopping amplitudes in Eq.(6) might be more difficult 
to handle. Therefore, Eq.(7) with all the $l$-dependence being in the form of a (possibly, complex)
on-site potential is likely to be better amenable to some analytic treatment. 

Previously, 
the pure $F_0$ model was shown to have no odd collective mode (shear-sound)
with the transverse current - but no density - oscillations. 
In fact, in the 3d FL there is a threshold for the minimal $F_1$ required to develop this mode.
By further including the higher $F_l$ one gives rise to the additional collective modes.
Moreover, the Coulomb interactions drastically modify the spectrum of longitudinal current and density fluctuations,
thus producing gapped plasmon modes while the transverse ones remain unaffected.

The earlier studies were either limited to the
collisionless limit ($\gamma_l\to 0$) \cite{else} 
or ignored the important cancellations (due to the conservation laws) between the quasiparticle 
self-energies $\Sigma$ and the  interaction function $F$ \cite{mandal1,mandal2}. 
Furthermore, the analyses of Refs.\cite{mandal1,else,mandal2} were oblivious to the
intrinsically peculiar 2d kinematics that tends to give rise to a strong disparity between 
the relaxation rates of the even and odd harmonics in the case of a convex (albeit not a concave)
FS \cite{tomo}. It has been extensively studied in the conventional Fermi liquid \cite{sarma,tomo} 
and also taken into account in the recent explorations of the 'Ising-nematic' variant of the problem (1) \cite{guo}.

The crux of the matter is that while the even harmonics can decay solely due to the FS shape fluctuations,  
the odd ones can not be relaxed without changes to the Fermi volume.
This kinematic property gives rise to a delayed onset of equilibration, thus 
resulting in the emergence of a new, 'tomographic', regime intertwining between
the ballistic (collisionless) and the ordinary hydrodynamic (diffusive) ones \cite{sarma,tomo}.  

In what follows, we consider a generic NFL system 
described by the singular momentum-independent self-energy $\Sigma(\omega)\sim\omega^{1/z}$ 
and the Landau function 
$
F(\theta |\omega)=F(q_{\parallel}/\omega^{1/z_{\parallel}}, q_{\perp}/\omega^{1/z_{\perp}})
$.
The latter favors small-angle scattering through the parametrically different scales of the 
normal and tangential
\be 
q_{\perp}\sim\omega^{1/z_{\perp}}~~~~~~ 
q_{\parallel}\sim\omega^{1/z_{\parallel}}
\ee
components of the transferred momentum, thus implying $q_{\parallel}/q_{\perp}<<1$ at low $\omega$,  
provided that $z_{\parallel}<z_{\perp}$. 

For instance, the Landau function corresponding to the interaction (1) takes the schematic form 
\be
F(\theta |\omega)\sim ({ max[|\theta|,\omega^{1/(\alpha+\beta)}]})^{d-2-\beta}
\ee
which results in the singular angular harmonics  
$
F_l(\omega)\sim\omega^{d-1-\beta/(\alpha+\beta)}
$
with $l\lesssim 1/\omega^{1/(\alpha+\beta)}$ for $d-1-\beta\leq 0$.

Alternatively, for a generalized bare quasiparticle 
dispersion 
$
\xi_p=Ap^{\eta}_{\parallel}+Bp^{\zeta}_{\perp}
$ 
the scaling of the normal and tangential FS fluctuations can be alternatively 
influenced by the sheer kinematics 
\bea  
q_{\perp}\sim (\omega+\Sigma)^{1/\zeta}\sim\omega^{1/z\zeta}
\nonumber\\
q_{\parallel}\sim(\omega+\Sigma)^{1/\eta}\sim\omega^{1/z\eta}
\eea
In the NFL regime where
$\Sigma>>\omega$ the estimates (8) and (10) would not necessarily be in agreement, thereby producing
different regimes and crossovers between them.  
However, the situation simplifies significantly if 
$z_{\perp}=z\zeta$ and $z_{\parallel}=z\eta$, thus implying
\be 
z_{\perp}/z_{\parallel}=\zeta/\eta
\ee
In the case of Eq.(1) this happens for all $\alpha=3-d$, including the physically important case of the 2d Landau-damped  
boson ($\alpha=1$) coupled to the fermions with an ordinary near-linear dispersion
($\eta=1$) and a quadratic curvature ($\zeta=2$).

Concomitantly, the two components of momentum scale differently with a characteristic scattering angle 
$
q_{\perp}\sim\theta
$
and
$
q_{\parallel}\sim\theta^{z_{\perp}/z_{\parallel}}
$, 
any power-law dependence upon which brings about 
a matching power of the angular momentum $l\sim 1/\theta$. 

As previously mentioned, 
the scattering processes with $\theta\lesssim\theta_0\sim\omega^{1/z_{\perp}}$ 
are dominated by the singular frequency dependence while for $\theta\gtrsim \theta_0$ all the scattering is 
essentially momentum dependent, hence quasi-elastic.
 
It is worth mentioning that the (un)conventional FL with $z\leq 1$ 
can also be fit into the above scaling laws, provided that instead of $\Sigma$ one takes  $\omega$  
as the greater of the two in Eq.(10).  

As regards the scattering rates, the 2d parity (equivalent to 1d 
reflection, $\theta\to - \theta$) and particle-hole ($\xi\to -\xi$) symmetries
require them to be proportional to the even powers of either, thus resulting in the following estimates
(here $n$ and $m$ are non-negative integers)  
\bea 
~~~~~~~~~~\gamma^{+}_l(\omega)\sim 
\Sigma(\omega) (q_{\perp}l)^{2n}
\sim \omega^{1/z+2n/z_{\perp}}l^{2n}~~~~~~~~~\nonumber\\
\gamma^{-}_l(\omega)\sim
\gamma^{+}_l(\omega)({q_{\parallel}}(l^2-1))^{2m}~~~~~~~~~~~\nonumber\\
\sim \omega^{1/z+2n/z_{\perp}+2m/z_{\parallel}} l^{2n}(l^2-1)^{2m}~~~~~~~~~~
\eea
for the even and odd angular harmonics, respectively. In accordance with the earlier discussions \cite{tomo}, the general 
structure of (12) accounts for the fact that the even-$l$ modes can be relaxed by 
the shape fluctuations of an incompressible FS (hence, factors of $q_{\perp}$) while the odd ones 
would also require changes to the FS volume (hence, factors of $q_{\parallel}$). 

Based on this argument, the minimal values of the integer factors in (12) are $n=0, m=1$. 
However, to keep the discussion as general as possible
and allow for an additional suppression of scattering through the matrix elements  
one might want to keep $n,m\geq 0$.
Conceivably, this additional suppression will affect the odd rates in the 
case of a convex FS, but not necessarily a concave one \cite{sarma,tomo}.

Despite being much lower at small $l$, the faster-growing 
odd rate catches up with the even one at the momenta of order 
$
l_{\parallel}=\omega^{-\eta/z_{\parallel}\zeta}
$, 
while at the still higher $l$ the two rates remain comparable. 

Also, at the momenta in excess of 
$
l_{\perp}=\omega^{-1/z_{\perp}}
$
the small-angle suppression in (12) ceases to exist.   
Thus, for $l>max[l_{\parallel},l_{\perp}]$
both the even and odd rates become of order the imaginary part of the self-energy $\Sigma$.
Also, as mentioned above, under the condition (11) one finds $l_{\perp}=l_{\parallel}\equiv l_0$
and the above crossovers merge into one and same.

In general, even a limited disparity between the even and odd rates  introduces new scales of order 
$(\gamma^{+}_{2l}\gamma^{-}_{2l+1})^{1/2}$ that mark transitions between some 
additional, different from, both, the ballistic 
and hydrodynamic, regimes.  

To apply (12) to the extensively studied case of the 2d FL with large-angle scattering and 
$\Sigma\sim \omega^2\ln\omega$ one has to choose $z=1/2$, $z_{\parallel}=1$, $z_{\perp}=\infty$,
$n=0$, and $m=1$, thus obtaining   
\be 
\gamma^{+}(\omega,l)\sim \omega^2~~~~~~~~~~
\nonumber\\
\gamma^{-}(\omega,l)\sim \omega^4(l^2-1)^2 
\ee
where $l>0$. 
Upon a closer inspection, the rates (13) acquire an additional $\ln l$ factor originating from the logarithmic
kinematic divergence in 2d \cite{tomo}.

By contrast, in the problem of the 
overdamped 2d bosonic mode (1) with $z=z_{\parallel}=3/2, z_{\perp}=3$ and $n=m=1$  
one readily recovers the results of Ref.\cite{guo} 
\be 
\gamma^{+}(\omega,l)\sim \omega^{4/3}l^2~~~~~~
\nonumber\\
\gamma^{-}(\omega,l)\sim \omega^{8/3}l^2(l^2-1)^2
\ee
In the kinetic equation (5), the imaginary terms are accompanied by their real-valued counterparts associated with the Landau function. Generically, they appear to be of the same order for any $z>1$.

Formally, both rates  $\gamma^{\pm}_{0}$ vanish due to the differences $F_0-F_l$ in Eqs.(6,7) which implement 
a cancellation between the quasiparticle self-energy and Landau function, as demanded by the particle number conservation.
Likewise, the vanishing of the odd rates  $\gamma^{-}_{\pm 1}$ is due to momentum conservation
and can be traced back to the factors $F_{\pm 1}-F_l=(F_0-F_l)-(F_0-F_{\pm 1})$ in the integrand of (5).\\

{\it Solving the kinetic equation}\\

Any essential $l$-dependence of either the on-site potentials or the on-bond factors in both
variants (6) and (7) of the kinetic equation  
precludes one from solving it analytically by deriving a closed formula for the 
infinite series $\sum_{l=0}\rho_lx^l$ \cite{sarma}, although a numerical solution might still be fairly 
straightforward.  

Likewise, a direct solution of the 1d Schrodinger equation is complicated by the fact that
the pertinent 1d potential oscillates strongly between the even and odd sites,
apart from the overall power-law growth of its envelop function up to the momenta
$l\sim l_0$. 
Despite the difficulty of finding even the classical turning points in such staggered potential,  
its leveling off at $l>l_0$ suggests that any potential discrete eigenstates  
would be confined to the range of momenta $|l|\lesssim  max[l_{\parallel}, l_{\perp}]$ where the effective 1d 
potential continues to grow with $|l|$.
At still higher harmonics the 1d potential levels off and its eigenvalues form a continuum 
of non-normalizable scattering states above the plateaux.
In contrast, collective zero-, shear-, and other sound modes correspond to the normalizable 
bound-state solutions, if any.

This picture is consistent with an  interpretation of the collective modes as soft FS oscillations with low $l$ 
that are delocalized over the FS while the particle-hole continuum represents rough local fluctuations
which include arbitrarily large $l$. 
To get a preliminary insight into the spectra of the 'strange sounds' 
one can first eliminate the even (fast) modes in favor of the odd (slow) ones 
in the kinetic equation, thereby deriving  
dispersion relations $\omega_l(q)$ for the different values of $l$.

Separating out the angular dependence, assuming (11), and extending the expressions (12)
for the rates (imaginary parts) to the full (complex-valued) bosonic 'self-energies' 
(for $z>1$ they would be, generally, of a comparable magnitude),
one then arrives at the schematic form of the spectral equation
\be 
(\omega+i\gamma^{+}_{2l}(\omega))(\omega+i\gamma^{-}_{2l+1}(\omega))=(vq)^2
\ee
Using this equation one can determine the exponent $\nu$ 
in the power-law dispersion relation 
$\omega(q)\sim q^{\nu}$. 

At low momenta, $l\lesssim 
l_+=\omega^{(1-1/z)/2n-1/z_{\perp}}
$, 
the self-energies can be neglected and the corresponding collective modes described by (15) exhibit the ordinary
linear dispersion, $\nu=1$. 

At the higher momenta, $l_{+} \lesssim l \lesssim l_{-}=\omega^{(1-1/z)/(2n+4m)-1/z_{\perp}}
$,  
the exponent in the algebraic dependence  dispersion changes to 
$ 
\nu={2/(1+1/z+2n/z_{\perp})}
$.

In the next interval, $l_-\lesssim l\lesssim l_0$, where $l_0$ was defined below Eq.(12)
the spectrum once again changes to 
$
\nu={1/(1/z+2n/z_{\perp}+m/z_{\parallel})}
$.

Finally, for $l>l_0$ the dispersion approaches the naive non-hydrodynamic asymptotic 
with
$ 
\nu=z
$
which was previously interpreted as marking the crossover between a particle-hole 'quasi-continuum' and (possibly) non-hydrodynamic modes \cite{qbe,else,mandal1,mandal2}.
The earlier studies differed, though, on whether the former would be composed of the modes lying 
below \cite{qbe,mandal1,mandal2} or above \cite{else} this divide.

The above analysis is approximate and, albeit yielding the anticipated results in the limits of, both, low and large $l$,
lacks accuracy in the intermediate regime. A further effort towards solving (5) with the kernal (9) would definitely be warranted.\\

{\it Response functions and transport rates}\\

A zero-sound mode would generally be observable as a peak in the (even)
density and  longitudinal current responses $<\rho |\rho>\sim<J_{\parallel}| J_{\parallel}>$
at the momenta where it separates from the particle-hole continuum. Likewise, the shear sound mode 
would manifest itself through the (odd) transverse current response function $<J_{\perp}| J_{\perp}>$. 

Generalized (possibly, off-diagonal) response functions $\chi_{AB}$ involving pairs of local fields
$|A>$ and $|B>$ are given by the general formula
\be 
\chi_{AB}=\sum_{O} <A|O>{1\over i\omega+\lambda_O}<O|B>
\ee
where the sum is taken over all the operators $|O>$ with eigenvalues $\lambda_O$ 
that have non-vanishing overlaps with the fields $|A>$ and $|B>$. 

In the case of electrical conductivity the current operator normally overlaps 
with momentum. If so, the so-called 'coherent' (Drude-like) term
in the conductivity remains finite only as long as momentum can relax. Alternatively, 
the current has to overlap with other conduction channels, as the result of which the conductivity may acquire some 
unconventional ('incoherent') contributions.

In the FL theory, the collective modes spectra can then be read off
directly from the poles of the fully dressed polarizabilities. 
Conceivably, this might still hold for the NFL, albeit in addition to (of instead of) the poles  
one might need to look into the possible branch cuts as well.

In particular, the singularities of a (generally, non-local) conductivity  
$
\sigma_{\perp,\parallel}(\omega,{\bf q})=\chi_{J_{\perp,\parallel}J_{\perp,\parallel}}/i\omega
$ 
represent contributions of all the modes ($l=0,\pm 1,\dots$) that have sizable overlaps 
with the current operator. 

While the rate $\gamma_1$ given by (12) 
vanishes for the isotropic FS, in any realistic situation it would be finite due to the combination of all
 the momentum non-conserving scattering processes (impurities, umklapp, phonons, etc.). 

Even in the absence of such processes, the odd rates with higher $l$ 
can still produce the optical ($q=0$) conductivity featuring a power-law tail at sufficiently high (yet, small compared to the Fermi energy) 
$\omega$  
\bea 
\sigma_{conv}(\omega,0)\sim Re \sum_{\pm,l}{|<J|O^{\pm}_l>|^2\over i\omega+ \gamma^{\pm}_l}
\nonumber\\
\sim \omega^{1/z+2n/z_{\perp}+2m/z_{\parallel}-2}
\eea
in the more restrictive case of a convex FS where $m\geq 1$ and  
$ 
\sigma_{conc}(\omega,0)
\sim \omega^{1/z+2n/z_{\perp}-2}
$
in the less restrictive concave one where the leading term is given by Eq.(12) with $m=0$. 

For instance, in the theory (1) it would result in   
the growing ($\sim\omega^{2/3}$) vs decaying ($\sim\omega^{-2/3}$) behavior
for the convex and concave FS, respectively, in agreement with the conclusions drawn in Refs.\cite{2/3}.
The former can then be recognized as the result of a cancellation of the leading terms 
by virtue of the vertex Ward identity due to momentum conservation.

At finite $q$, by keeping only the lowest $\gamma^{-}_1$ and $\gamma^{+}_2$ rates
and equating the denominator in (17) to zero 
one finds the approximate dispersion relations  
\be 
\omega^{\pm}={i\over 2}(\gamma^{+}_{2}+\gamma^{-}_{1})\pm 
{i\over 2}[(\gamma^{+}_{2}-\gamma^{-}_{1})^2-4(vq)^2]^{1/2}
\ee
which evolve from the ballistic linear dispersion at large $q$ all the way down to the viscous one
$\omega=-iq^2/\gamma^{+}_{2}$  at small $q$ and $\gamma^{-}_1\to 0$.

Notably, the strong $\omega$-dependence of the rates $\gamma^{\pm}_l$ results in an unusual viscous behavior,
as compared to the conventional hydrodynamic (Stokes) one (cf. \cite{tomo}). 

To further elaborate on this topic and better assess the possible departures from the standard hydrodynamics 
one can also try to evaluate the eigenvalues $\lambda_O$ 
by accommodating the slow angular diffusion directly in the effective 
kinetic equation for the odd-parity fluctuations $\rho^{-}(\theta)=-\rho^{-}(-\theta)$.  
This equation can be obtained by integrating out the parity-even modes $\rho^{+}(\theta)=\rho^{+}(-\theta)$
which procedure results in the highly non-local (super)diffusion equation generalizing that derived in Refs.\cite{tomo}
\be 
(i\omega+{(vq)^2\over \gamma^{+}}
\theta^2{\partial_{\theta}^{-2n}}
+
\gamma^{-}\partial_{\theta}^{2n+4m})\rho^{-}(\theta)=0
\ee
The minimal eigenvalue of this operator can be estimated 
by minimizing the l.h.s.of (19) at  the angle 
$ 
\theta_{min}\sim ({\gamma^{+}\gamma^{-}/v^2q^2})^{1/(2+4n+4m)}
$.

Estimated at this value, the conductivity exhibits a characteristic
 frequency $\omega(q)$ whose scaling with $q$ yields a new exponent 
\be 
\nu={2n+4m\over 1+2n+2m+({1\over z}+{2n\over z_{\perp}})(2m-1)-{2m(1+n)\over z_{\parallel}}}
\ee
signifying the unconventional hydrodynamics.  

To further improve on the estimate (20) and account for the $l$-dependence in Eq.(12)  
one can also evaluate the specific conductivity-related decay rate \cite{tomo}. 
In this approach, the (non-local) conductivity 
\be 
\sigma(\omega,q)\sim {1\over i\omega+q^2/\Gamma_2(\omega,q)}
\ee
is expressed in terms of the effective rate $\Gamma_2(\omega,q)$ 
which is itself is given by the $l=2$ value of the continued fraction
\be 
\Gamma_l=\gamma_l+
{(vq)^2\over {\gamma_{l+1}+{(vq)^2\over {\gamma_{l+2}+\dots}}}}
\ee
 that satisfies the finite-difference equation  
\be 
(\gamma_l-\Gamma_l+\Gamma_{l+2})((vq)^2+\gamma_{l+1}\Gamma_{l+2})=\gamma_{l+1}\Gamma^2_{l+2}
\ee 
In the continuum limit, this equation can be converted into a differential one. 
Namely, by virtue of a smart parametrization \cite{tomo} 
\be 
\Gamma_l={(vq)^2\over \gamma_{l-1}}
({\psi_l\over \psi_{l+2}}-1)\approx -2{(vq)^2\over \gamma^{-}}{d\ln\psi(l)\over dl}
\ee
the function $\Gamma_l$ can be cast in terms of a solution $\psi(l)$ to the equation 
\be 
{d^2\psi\over dl^2}-{1\over \gamma_{-}}{d\gamma_{-}\over dl}{d\psi\over dl}-{\gamma_{+}\gamma_{-}\over 4(vq)^2}\psi=0
\ee
which take the form
\be 
\psi(l)\sim x^{\mu}K_{\mu}(x)
\ee
where
$
\mu=(1+2n+4m)/(2+4n+4m)
$ 
while $K_{\mu}(x)$ is the modified Bessel function of the second kind, and 
$
x=({\gamma_{+}\gamma_{-})^{1/2}l^{1+2n+2m}/2(vq)(1+2n+2m)} 
$ 
is its dimensionless argument.
 
Thus, the rate of interest shows a significant spatio-temporal dispersion
\be
\Gamma_2(\omega,q)\sim
(vq)^{2-2\mu}{(\gamma_{+})^{\mu}
\over (\gamma_{-})^{1-\mu}}+{\#}\gamma_{+}
\ee
where the q-dependent term dominates over the constant in the range  
of momenta $(\gamma_{-}\gamma_{+})^{1/2}\lesssim q\lesssim \gamma_{+}$
which defines the intermediate 'tomographic' regime.

In the NFL regime $\Sigma>>\omega$ Eq.(27) reads 
\be
\Gamma_2(\omega,q)\sim (q^{1+2n}\omega^{{2m\over z}+{4nm\over z_{\perp}}-{m(1+2n)\over z_{\parallel}}})^{1/(1+2n+2m)}
\ee
The position of the pole in (21) then yields yet another exponent 
\be
\nu={1+2n\over 1+2n+2m+2m/z+4nm/z_{\perp}-m(1+2n)/z_{\parallel}}
\ee
For instance, in the 2d FL with $z=2=2z_{\parallel}$ and $z_{\perp}=\infty$ the rate is  
$\Gamma_2\sim q^{1/3}\omega$ \cite{tomo}.
In the universality class of a 2d Ising nematic with $z=3/2=z_{\parallel}=z_{\perp}/2$ one obtains   
$\Gamma_2\sim q^{3/5}\omega^{2/15}$, as in Ref.\cite{guo}.
 
Notably, the above exponents 
appear to be consistently smaller and, therefore, more relevant than the larger ones 
controlling the quasiparticle self-energies.
Even in the FL case the rate (27) turns out to be 
linear in temperature - despite the intrinsically non-linear rates (12) 
 - thus creating the appearance of a NFL behavior in those experiments that 
measure the non-local conductivity \cite{tomo}.

Thus, the damping rate $\Gamma_2$ manifested by the non-local conductivity and associated with kinematic viscosity 
turns out to be larger than its counterparts 
in, both, the ballistic (linear) regime, as well as the viscus (quadratic) one. 
In light of this observation one might want to revisit, both, the issue of relative (un)importance
 of the higher-order hydrodynamic corrections as well as that of the (ostensibly) NFL-like effects in realistic systems.  

On a cautionary note, the previous work on the tomographic regime in FL 
dealt exclusively with the well-defined temperature-dependent rates. However,
at finite temperatures the interaction (1) gives rise to the strongly enhanced (potentially divergent) 
thermal fluctuations at the zero Matsubara frequency, unless the overdamped boson 
develops a finite 'thermal mass' which regularizes the pertinent infrared divergence \cite{gaugeold}.
While this formal problem remains lingering in the case of an unbroken gauge symmetry, one might surmise
that it can be overcome by operating in terms of manifestly 
gauge-invariant fermion amplitudes instead of the ordinary (gauge non-invariant) ones \cite{rmf}.\\

{\it Betterment of the kinetic equation}\\

Prior to the recent revival of interest in the firmly established conservative approaches, such as the FL theory and kinetic equation, the problem of compressible strange metals was also attacked with such 'experimental' 
techniques as multi-dimensional bosonization and applied holography. 

As regards the former, apart from the straightforward attempts to discretize the FS into a collection of (pseudo)-1d 'patches' \cite{bosold} (which method would be intrinsically plagued with the lack of a proper account of the FS curvature \cite{chubdvk}), there was also some (apparently, premature) effort made towards formulating a purely geometrical approach to bosonization in the framework of the Kostant-Kirillov method of coadjoint orbit quantization \cite{geom}. 

Recently, the latter approach was re-discovered anew \cite{bosnew} - incidentally, 
after it had been invoked in the context of a potentially even more fundamental approach of 
a path integral over the Wigner distribution function-like variables defined in the phase space \cite{hydro}. 

Thus far, though, the recent expositions of this method 
\cite{bosnew} were largely limited to the discussions of the mathematical aspects of the formalism itself, straightforward RPA-like analyses, 
and multi-point correlations of the free density operators. 
Such correlations are controlled by the fermion kinematics 
and stem from the cubic and quartic terms in the 
effective free-fermion bosonic action due to its intrinsic 
non-linearity even in the absence of any physical interactions.

At the core of all the different variants of the bosonization method, is the (approximately) quadratic bosonic action   
\be
S=i\int_{r,t}\oint_{\bf n}{\bf n}{\bf \nabla}\phi_{\bf n}({\partial\over \partial t}-v{\bf n}{\bf \nabla})\phi_{\bf n}
\ee
for the field $\phi$ representing density fluctuations via $\rho(t,r)=\oint_{\bf n}{\bf n}{\bf \nabla}\phi_{\bf n}$
and governed by the propagator
\bea
<\phi_{n}\phi_{n^{\prime}}>
={\delta({\bf n}-{\bf n}^{\prime})\over ({\bf n}{\bf q})(\omega-v({\bf n}{\bf q})-\Sigma)}+\nonumber\\
{F({\bf n},{\bf n}^{\prime})\over 
(\omega-v({\bf n}{\bf q})-\Sigma)(\omega-
v({\bf n}^{\prime}{\bf q})-\Sigma)}
\eea
The task of computing this function is essentially equivalent to inverting the collision operator
or, for that matter, solving the linearized kinetic equation (5).

When applied to the interaction (1) it results in the equations  
\be 
(\omega-v({\bf n}{\bf q})+A\omega^{2/3})\phi^{\parallel,\perp}_{\bf n}=B\omega^{2/3}
\oint_{{\bf n}^{\prime}}({\bf n}_{\parallel,\perp}{\bf n}^{\prime}_{\parallel,\perp})\phi^{\parallel,\perp}_{{\bf n}^{\prime}}
\ee
which were used in the early exploration of this problem to
produce the anomalous dispersion relation $\omega\sim k^{6/5}$ (see the second of Refs.\cite{stampdvk})
which was recently reproduced in \cite{mandal2}, in addition to the obvious one, $\omega\sim k^{3/2}$.

While the latter was identified early on as 
marking a crossover between the particle-hole continuum 
and possible non-hydrodynamic modes \cite{qbe}, the former mode was argued  
to evolve into the ordinary linear zero-sound at higher momenta \cite{mandal2}.
Notably, this mode was found in the parity-even - but not the parity-odd - sector.

In order to further improve the above estimates and better ascertain the nature of the modes in question,
it might be possible to advance the bosonization calculations by advancing the eikonal technique of 
the early Refs.\cite{stampdvk}. Apart from the low-$q$ regime, this method can also 
be used to compute the response functions at the transferred momenta close to $2k_F$.

The hydrodynamics of strange metals was also actively pursued in the framework of the 
so-called applied holography \cite{ads}.
By now this speculative (a.k.a. 'bottom-up') bulk-boundary correspondence 'which has proven in recent years to 
effectively describe low-energy properties of strongly interacting systems' (as 
quoted in, e.g., the last of Refs.\cite{plasmon}) has already withdrawn from much of the territory in 
condensed matter physics that it held since 2007.
Nowadays it focuses almost exclusively on (and claims to provide new insights into) the various hydrodynamic 
aspects of strongly (or weakly, as in a typical holographic calculation
the coupling strength is not even a factor) interacting matter. 

Among its many broadly publicized (yet, either unconfirmed or refutable \cite{mystery}) propositions, there
is a recent - rather specific and purportedly quantitative - prediction of an overdamped acoustic plasmon mode and its properties in the double-layered cuprates  \cite{plasmon}.
Similar to the overwhelming majority of other calculations in 'orthodox' holography this one was performed 
far outside the regime of applicability of the underlying theory, thus raising questions about the status of any quantitative agreement with data, should such be found (a spoiler: unless such an agreement is deemed purely fortuitous, 
that theory would probably have much more explaining to do, as compared to a situation of no agreement).  

Specifically, 
apart from the generally missing translational, rotational, and/or super-symmetry 
of a large number of fermion species (which would be of order unity in any metal - strange or not),  
the physically tangible interactions normally do not happen to be 
in the regime of extremely strong couplings that would be required to potentially justify the purely
classical treatment of the bulk gravity. Besides, the ubiquitous holographic custom of picking out 
that bulk theory at will (or on the basis of such 'anthropic' factors as technical convenience and prior insight) 
and regardless of the material in question does not help the holographic cause either \cite{mystery}.
 
The prediction of Refs.\cite{plasmon} was argued to be in conflict with
the earlier resonant inelastic X-ray scattering (RIXS) \cite{RIXS} and
Electron Energy Loss Spectroscopy (EELS) \cite{abba}
studies which detect a well-defined acoustic plasmon at small momenta, followed by a transition into a featureless 
momentum-independent continuum at larger momenta. 

In accord with such conclusions, the most recent EELS studies report that
'there are no signs of over-damped plasmons predicted by holographic theories',
'the theoretical predictions of an over-damped plasmon are in stark conflict with 
early EELS studies on cuprates'
or 'the plasmons which are calculated by theory are unlikely candidates to explain the dispersion curves'
(see the first and second of Refs.\cite{noplasmon}, respectively). 

In that regard, it might be instructive to quote the work of the last of Refs.\cite{plasmon} for a rare example of 
unequivocal self-confidence (albeit somewhat less of a healthy self-criticism): 
'Our theoretical predictions appear to contradict the EELS 
results which are obtained by our experimental research group, in which no acoustic 
plasmon was observed [see \cite{noplasmon}]. This apparent contradiction certainly challenges the experimentalist to 
either discover the right experimental conditions to observe the acoustic plasmon contribution
or come up with arguments to explain why an acoustic plasmon cannot be measured.'

As to a potential relevance of the holographic claims towards hydrodynamics,  
such predictions tend to be limited to the linear (transformed into a square-root in the presence of
the unscreened Coulomb interactions) and quadratic modes for the longitudinal 
(sound) and transverse (shear) modes, respectively. Therefore,
reproducing any of the above (in general, non-analytical) dispersion relations 
as quasi-normal modes in a certain gravitational background metric 
would pose an interesting challenge to the 'orthodox' holography. In contrast to the situation in higher dimensions,  reconciling the results of solving the kinetic eqiation
with those based on the $2d$ hydrodynamics can be particularly involved because of the  
presence of the intermediate 'tomographic' regime.  
Likewise, it would be interesting to ascertain 
a possible role of the much-discussed holographic phenomenon of 'pole-skipping' 
in the context of the kinetic equation and its improved versions.  

On the other hand, one might find some intriguing hints of a conformally symmetric behavior (hence, some potential of
an underlying gravitational physics) because of the factor $l(l^2-1)$ in (12) which is inconspicuously  
reminiscent of the Virasoro commutations relations on the algebra of 1d diffeomorphisms. 

As another tantalizing observation, in the 'tomographic' regime 
the kinetic equation (5) can formally be converted to the 1d 
quantum mechanics of a particle in the solvable (supersymmetric) potential $f_0(1-f_0)\sim 1/\cosh^2x$ \cite{tomo}, 
which simple problem played an important role in the recent studies of 
the Sachdev-Ye-Kitaev (SYK) \cite{syk} and related ultra-local models that can be viewed as 
low-dimensional examples of the 'Hall-ographic' bulk-boundary correspondence \cite{Hall}. 

Some difference, though, is that in terms of the two-point fermion amplitude (3) 
the SYK quantum mechanics emerges along the direction of the relative (thermal) time variable $\tau$,
whereas in the kinetic equation the role of the 
1d coordinate $x$ is played by the (dimensionless) energy $\epsilon/T$ dual to the center-of-mass time variable.

Possible rationalization of these and related observations in the framework of the 'phase-space holography' 
introduced in Ref.\cite{hydro} will be presented elsewhere.\\

{\it Reflections and repercussions}\\ 

The above discussion suggests that the question of the existence of well-defined collective modes ('sounds') 
in strongly interacting ('strange') metals may lie outside the standard FL paradigm 
of stable (undamped) bound states positioned above the particle-hole continuum.  
It was observed that, while remaining nearly linear  (as long as the Coulomb interactions are excluded)
and relatively weakly damped at large momenta, 
such modes tend to acquire non-linear dispersion and strong damping at lower momenta. 

Moreover, such transport characteristics as a non-local conductivity might exhibit certain NFL-like
features that would be routinely associated with the 'strange metals'. Even such an alleged 
hallmark of the NFL behavior as linear resistivity was shown in \cite{tomo} to occur even in the ordinary FL
transport, if measured in a certain regime.
    
As proposed earlier \cite{chubmas}, the poles and branch cuts of the charge/current response functions 
 manifest themselves through their distinct behavior in the time domain, 
$
\chi(t,{\bf q})\sim e^{-i\omega(q)t}
$,
 which might be accessible with the experimental pump-probe techniques.
In particular, one can discuss specific features associated with
the detectable vs hidden (or 'mirage') poles of the charge susceptibility \cite{chubmas}.

Intriguingly, in Ref.\cite{guo} it was conjectured that the negative real part of the analytically continued odd rate (12) 
may signal a potential instability of the system with a convex FS. While being in line with the general argument, this conclusion would seem rather puzzling  
as, under the same conditions, the much greater even rate would not indicate any instability.    
Therefore, it would seem unlikely that the less relevant (containing a higher power of $\omega$) 
rate would be indicative of an instability while the more relevant (proportional a lower power) does not. 

These and related issues are definitely worth pursuing.

\end{document}